\documentclass[runningheads]{llncs}
\def\BibTeX{{\rm B\kern-.05em{\sc i\kern-.025em b}\kern-.08em
    T\kern-.1667em\lower.7ex\hbox{E}\kern-.125emX}}
\usepackage{graphicx}
\usepackage{mathtools}
\usepackage{lipsum} 
\usepackage[numbers]{natbib}

\usepackage{booktabs}
\usepackage[ruled]{algorithm2e}
\usepackage{url}
\usepackage{color}
\usepackage{multirow}
\usepackage{enumitem}
\usepackage{array}
\usepackage{siunitx}
\usepackage[labelfont={bf}]{caption}
\usepackage{balance}
\usepackage{array} 
\usepackage{todonotes}
\usepackage{subfig}
\usepackage[nolist,nohyperlinks]{acronym}
\usepackage{footnote}
\usepackage[perpage]{footmisc}
\usepackage[english]{babel}
\usepackage[autostyle]{csquotes}
\makesavenoteenv{tabular}

\acrodef{DL}{Deep learning}
\acrodef{CNN}{Convolutional Neural Network}
\acrodef{MIS}{Minimally Invasive Surgery}
\acrodef{AI}{Artificial Intelligence} 
\acrodef{CAI}{Computer Assisted Intervention} 
\acrodef{ML}{Machine Learning}
\acrodef{ReLU}{Rectified linear Unit}
\acrodef{mIoU}{mean Intersection over Union}
\acrodef{AUC}{Area Under Curve}
\acrodef{ROI}{Region of Interest}
\acrodef{MICCAI}{Medical Image Computing and Computer Assisted Intervention Society}
\acrodef{IOU}{Intersection over Union}
\acrodef{DSC}{Dice coefficient}
\acrodef{GI}{gastrointestinal}
\acrodef{ASPP}{Atrous spatial pyramid pooling}

\begin{document}
% \title{Kvasir-Instrument: An instrument segmentation dataset in gastrointestinal endoscopy}
\title{Kvasir-Instrument: Diagnostic and therapeutic tool segmentation dataset in gastrointestinal endoscopy}

\author{Blinded}
\authorrunning{Blinded et al.}
\institute{Blinded}

\author{Debesh Jha\inst{1,2}\and 
Sharib Ali\inst{9}\and 
Krister Emanuelsen \inst{3}\and
Steven A. Hicks\inst{1,5}\and 
Vajira Thambawita\inst{1,5} \and 
Enrique Garcia-Ceja\inst{10}\and 
Michael A. Riegler\inst{1}\and \\
Thomas de Lange\inst{4,6,7}\and
Peter T. Schmidt\inst{8}\and
H\aa vard D. Johansen\inst{2}\and \\
Dag Johansen\inst{2}\and
P\aa l Halvorsen\inst{1,5}
}

\authorrunning{Jha et al.}
%\authorrunning{Blinded et al.}

\institute{SimulaMet, Norway \and UIT The Arctic University of Norway \and Simula Research Laboratory, Norway \and  Augere Medical AS, Norway \and Oslo Metropolitan University, Norway \and Medical Department, Sahlgrenska University Hospital-Mölndal, Sweden \and Department of Medical Research, Bærum Hospital, Norway \and Karolinska University Hospital, Sweden \and Dept. of Engineering Science, University of Oxford, Oxford, UK  \and Sintef Digital, Norway\\
\email{debesh@simula.no}
}

\maketitle            
\begin{abstract}
Gastrointestinal (GI) pathologies are periodically screened, biopsied, and resected using surgical tools. Usually the procedures and the treated or resected areas are not specifically tracked or analysed during or after colonoscopies. Information regarding disease borders, development and amount and size of the resected area get lost.
This can lead to poor follow-up and bothersome reassessment difficulties post-treatment. To improve the current standard and also to foster more research on the topic we have released the  ``Kvasir-Instrument'' dataset which consists of $590$ annotated frames containing GI procedure tools such as snares, balloons and biopsy forceps, etc. Beside of the images, the dataset includes ground truth masks and bounding boxes and has been verified by two expert GI endoscopists. Additionally, we provide a baseline for the segmentation of the GI tools to promote research and algorithm development. We obtained a dice coefficient score of $0.9158$ and a Jaccard index of $0.8578$ using a classical U-Net architecture. A similar dice coefficient score was observed for DoubleUNet. The qualitative results showed that the model did not work for the images with specularity and the frames with multiple instruments, while the best result for both methods was observed on all other types of images. Both, qualitative and quantitative results show that the model performs reasonably good, but there is a large potential for further improvements. 
Benchmarking using the dataset provides an opportunity for researchers to contribute to the field of automatic endoscopic diagnostic and therapeutic tool segmentation for \ac{GI} endoscopy.
\end{abstract}

\section{Introduction}
\label{sec:introduction}
\acf{MIS} is a commonly used technique in surgical procedures. The advantage of \ac{MIS} is that small surgical incisions are made in the patient for endoscopy that causes less pain, reduced time of the hospital stay, fast recovery, reduced blood loss, and less scaring process as compared to the traditional open surgery. The nature of the operation is complex, and the surgeons have to precisely tackle hand-eye coordination, which may lead to restricted mobility and a narrow field of view~\cite{bernhardt2017status}. 

However, unlike the treatment of accessory organs such as liver and pancreas, no incision is required for \ac{GI} tract organs (\textit{oesophagus, stomach, duodenum, colon, and rectum}). \ac{GI} procedures also includes both, minimally invasive surveillance and treatment (\textit{including surgery}) procedures. A varied number of tools are used as per the requirement of these procedures. For example, balloon dilatation to help open the \ac{GI} surface, biopsy forceps for tissue sample collection, polyp removal with snares and submucosal injections.

A computer and robotic-assisted surgical system can enhance the capability of the surgeons~\cite{cleary2010image}. It can provide the opportunity to gain additional information about the patient, which can be useful for decision making during surgery~\cite{bodenstedt2018comparative}. However, it is difficult to understand the spatial relationship between surgical instruments, cameras, and anatomy for the patient~\cite{pakhomov2019deep}. In \ac{GI} tract endoscopy, it is vital to track and guide surgeons during tumor resection or biopsy collection from a defined site, and help to correlate the biopsied samples and treatment locations post-diagnostic and therapeutic or surgical procedures. While most datasets and automated-algorithm developments for instrument segmentation are mostly focused on laparoscopy-based surgical removal, automatic guidance of tools for \ac{GI} tract surgery has not been addressed before.
%

% pictures from polyp removal with snares, balloon dilatations, submucosal injections and some other procedures.   
New developments in the area of robot-assisted systems show that there is potential for developing a fully automated robotic surgeon~\cite{shvets2018automatic}. The da Vinci robot is a surgical system that is considered the de-facto standard-of-care for certain urological, gynecological, and general procedures~\cite{allan20192017}. Thus, it is critical to have information regarding the intra-operative guidance, which plays an essential role in decision making. However, there are specific challenges, such as limited field of view and difficulties with the surgeons handling the instruments during surgery~\cite{ross2020robust}. Therefore, image-based instrument segmentation and tracking are gaining more and more attention in both robotic and non-robotic minimally invasive surgery. Previous work targeting instrument segmentation, detection, and tracking on endoscopic video images failed on challenging images such as images with blood, smoke, and motion artifacts~\cite{ross2020robust}. Other reasons that make semantic segmentation of surgical instruments a challenging task are the presence of images containing shadows, specular reflections, blood, camera lens fogging, and the complex background tissue~\cite{shvets2018automatic}. The segmentation masks of these images can be useful for instrument detection and tracking. 
% This part specific focus to GI tract
Similarly, in the \ac{GI} tract procedures, from tissue sample collection to surgical removal of pathologies is performed in low field-of-view areas. Visual clutter such as artifacts, moving objects, and fluid, hinders the localisation of the target site during surgical procedures. Additionally, currently, there is no way of correlating the tissue sample collection with biopsied location and assessing surgical procedure effectiveness or even post-treatment recovery analysis. Automated localisation and tracking of instruments can help guide the endoscopists and surgeons to perform their tasks more effectively. Also, post-procedure video analysis can be done using these automated methods to track such tools, thus enabling improved surgical procedures or surveillance and their post-assessment. Currently, this is an open problem in the research community, where most procedures are not automated in \ac{GI} tract endoscopy. 

While there is an open research question for the automated tool detection and guidance in GI procedures, there is a lack of available public datasets. We aim to initiate the development of automated systems for the segmentation of \ac{GI} tract diagnostic and therapeutic endoscopy tools. This research direction will enable tracking and localisation of essential tools used in endoscopy and help to improve targeted biopsies and surgeries in complex GI tract organs. To accomplish this, and to address the lack of publicly available labeled datasets, we have publicly released $590$ pixel-level annotated frames that comprise of tools such as balloon dilation for facilitating opening of \ac{GI} organs, biopsy forceps for tissue sample collection, polyp removal with snares, submucosal injections, radio-frequency ablation of dysplastic mucosa using probes and some other related surgical/diagnostic procedures. The released video frames will allow for building automated machine learning algorithms that can be applied during clinical procedures or post-analyses. To commence this effort, we provide a baseline benchmark on this dataset. U-Net~\cite{ronneberger2015u} is a common semantic segmentation based architecture for medical image segmentation tasks. In this paper, we thus present results utilising two U-Net based architectures. The provided dataset is open and can be used for research and development, and we invite multimedia researchers to improve over the provided baseline methods. The main contributions of this paper are:

\begin{itemize}
\item Release of $590$ annotated bounding box and segmentation masks of GI diagnostic and surgical tool dataset. To the best of our knowledge, this is the first dataset of segmented tools in the GI tract.

\item Benchmark of the provided dataset using U-Net and DoubleUNet architectures for semantic segmentation. Standard computer vision metrics are used for a fair comparison of methods and possible future work. 
\end{itemize}

%Section~\ref{sec:related work} presents the related work. Section~\ref{sec:Material} presents the method used in work. In Section~\ref{sec:Results}, we show the results of the proposed methods. In Section~\ref{sec:discussion}, we show the discussion. In Section~\ref{sec:conclusion}, wee conclude our work and provide the future directions. 

\begin{table}[!t]
\footnotesize
\def\arraystretch{1.1}
\caption{Available instrument datasets}
\label{tab:availabledataset}
\begin{tabular}{|l|l|l|l|}
\toprule
\textbf{Dataset} & \textbf{Content} & \textbf{Task  type} & \textbf{Procedure} \\ \midrule
\begin{tabular}[c]{@{}l@{}}Instrument \\ segmentation\\ and tracking (2015)~\cite{bodenstedt2018comparative}\end{tabular} & \begin{tabular}[c]{@{}l@{}}Rigid and robotic\\  instruments\end{tabular} & \begin{tabular}[c]{@{}l@{}}Segmentation \\ and tracking\end{tabular} & Laparoscopy\\ \hline
\begin{tabular}[c]{@{}l@{}}Robotic Instrument \\ Segmentation (2017)~\cite{allan20192017}\end{tabular} & \begin{tabular}[c]{@{}l@{}}Robotic surgical\\  instruments\end{tabular} & \begin{tabular}[c]{@{}l@{}}Binary segmentation,\\ part based \\ segmentation, \\instrument\\ segmentation\end{tabular} & \begin{tabular}[c]{@{}l@{}}Abdominal\\ porcine \end{tabular}\\ \hline

\begin{tabular}[c]{@{}l@{}}Robotic Scene\\ Segmentation (2018)~\cite{allan20192018}\end{tabular} & \begin{tabular}[c]{@{}l@{}}Surgical instruments\\ and other  \end{tabular}& \begin{tabular}[c]{@{}l@{}}Multi-instance\\  segmentation\end{tabular} & \begin{tabular}[c]{@{}l@{}}Robotic\\ nephrectomy  \end{tabular}\\ \hline

\begin{tabular}[c]{@{}l@{}}Robust Medical\\ instrument\\ segmentation (2019)~\cite{ross2020robust}\end{tabular} & \begin{tabular}[c]{@{}l@{}} laparoscopic\\ instrument \end{tabular} & \begin{tabular}[c]{@{}l@{}}Binary  segmentation, \\ multiple instance\\ detection,  multiple  \\ instance segmentation\end{tabular} & Laparoscopy \\ \hline
Kvasir-Instrument & \begin{tabular}[c]{@{}l@{}} Diagnostic and\\ therapeutic tools\\ in endoscopic images\end{tabular} & \begin{tabular}[c]{@{}l@{}}Binary segmentation\\ Detection and\\ localization\end{tabular} & \begin{tabular}[c]{@{}l@{}}Gastroscopy \\ \& colonoscopy \end{tabular}\\ \bottomrule
\end{tabular}
\end{table}

\section{Related Work}
\label{sec:related work}
Surgical vision is evolving as a promising technique to segment and track instruments using endoscopic images~\cite{bodenstedt2018comparative}. To gather researchers on a single platform, \textit{Endoscopic vision (EndoVis) challenge} is being organized since 2015 at \ac{MICCAI} with an exception in 2016. The Endovis challenge hosts different sub-challenges. The year-wise information about the hosted sub-challenge can be found on the challenge website\footnote[1]{\url{https://endovis.grand-challenge.org/}}.

Bodenstedt et al.~\cite{bodenstedt2018comparative} organized "EndoVis 2015 Instrument sub-challenge" for developing new techniques and benchmarking the \ac{ML} algorithm for segmentation and tracking of the instruments on a common dataset. The organizers challenged on two different tasks, (1) Segmentation, (2) Tracking. The goal of the challenge was to address the problem related to segmentation and tracking of articulated instruments in both laparoscopic and robotic surgery\footnote[2]{\url{https://endovissub-instrument.grand-challenge.org/EndoVisSub-Instrument/}}. A comprehensive evaluation of the methods used in instrument segmentation and tracking task for minimally invasive surgery is summarized in this work~\cite{bodenstedt2018comparative}. The extensive evaluation showed that deep learning works well for instrument segmentation and tracking tasks. 

In 2017, a follow up to the previous 2015 challenge was organized called "Robotic Instrument Segmentation Sub-Challenge"\footnote[3]{\url{https://endovissub2017-roboticinstrumentsegmentation.grand-challenge.org/}}. The challenge was part of the Endoscopic vision challenge that was organized at \ac{MICCAI} 2017. This challenge offered three tasks: (1) Binary segmentation task, (2) Parts based segmentation task, and (3) Instrument type segmentation task. The goal of the binary segmentation task was to separate the image into an instrument and background. Parts segmentation challenged the participants to divide the binary instrument into a shaft, wrist, and jaws. Type segmentation challenged the participants to identify different instrument types. A detailed description of the challenge tasks, dataset, methodologies used by ten participating teams in different tasks, challenge design, and limitation of the challenge can be found in the challenge summary paper~\cite{allan20192017}. 

In 2019, a similar challenge called "Robust Medical Instrument Segmentation Challenge 2019"\footnote[4]{\url{https://robustmis2019.grand-challenge.org/}} was organized by Roß et al.~\cite{ross2020robust}. This challenge offered three tasks (1) Binary segmentation, (2) Multiple instance detection, and (3) Multiple instance segmentation. The challenge was focused on addressing two key issues in surgical instruments, \textit{Robustness} and \textit{Generalization}, and benchmark medical instrument segmentation and detection on the provided surgical instrument dataset. EAD2019 challenge focused on endoscopic artefact detection primarily, but also included instrument class in their detection, segmentation and ``out-of-sample'' generalisation tasks. The challenge outcome revealed that most methods performed well for instrument detection and segmentation class~\cite{ali2020objective}. However, this dataset mostly consisted of large biopsy forceps. 

%In addition to these challenges, there are works on instrument segmentation in robot-assisted surgery~\cite{shvets2018automatic,pakhomov2019deep}, neurosurgical instrument segmentation~\cite{kalavakonda2019autonomous}, surgical instrument segmentation on mimimally-invasive robot-assisted procedures~\cite{azqueta2019segmentation}, instrument tracking~\cite{laina2017concurrent}, endoscopic instrument segmentation~\cite{isensee2020or}, and instrument recognition in laparoscopy~\cite{kletz2020instrument}. 

In Table \ref{tab:availabledataset}, we present available instrument datasets in the field. All of the datasets were designed for hosting challenges. The training dataset is released for all the datasets (except ROBUST-MIS); however, the test dataset is not provided by the challenge organizers. Thus, it makes it difficult to calculate and compare the results on the test dataset. However,  experiments are still possible by splitting the training dataset into train, validation, and testing sets. The Robust Medical instrument segmentation dataset is yet not public. However, the participants who have participated in the challenge have the opportunity to download the training dataset. Usually, there are certain practicalities to download the dataset, such as signing the agreement and, getting permission from the owner, which takes time, and it is inconvenient.
Moreover, to participate in the challenge, the participants have to signup in the particular year, and usually, the organizers do not make the dataset public unless they make a publication out of it, meaning it may take up to years. Thus, the significance of the datasets becomes less as the technology is changing rapidly. More information on available instrument datasets, contents, and offered tasks by the organizers and about the availability can be found from Table~\ref{tab:availabledataset}. 

The literature review shows that there are only a few open-access datasets for \ac{MIS} instrument segmentation. However, to the best of our knowledge, \ac{GI} tract organ tools have never been explored. This is the first attempt to identify this avenue and provide the community with a curated and annotated public dataset that comprises of diagnostic and therapeutic tools in the GI tract. We believe that the presented dataset and the widely used U-Net based algorithm benchmark will encourage the multimedia researchers to develop a robust and efficient algorithm on the provided dataset that can help clinical procedures in endoscopy.  
\section{Kvasir-Instrument dataset}
\label{sec:Material}
% TODO: 
% Example images
% Analysis of dataset 
% Representative examples from dataset (showing diversity)
% TODO: Debesh--> compute evaluation table and pull top and worse result samples (qualitative)
%
\begin{figure}[t!]
    \centering
    \includegraphics[width= 1.0\textwidth]{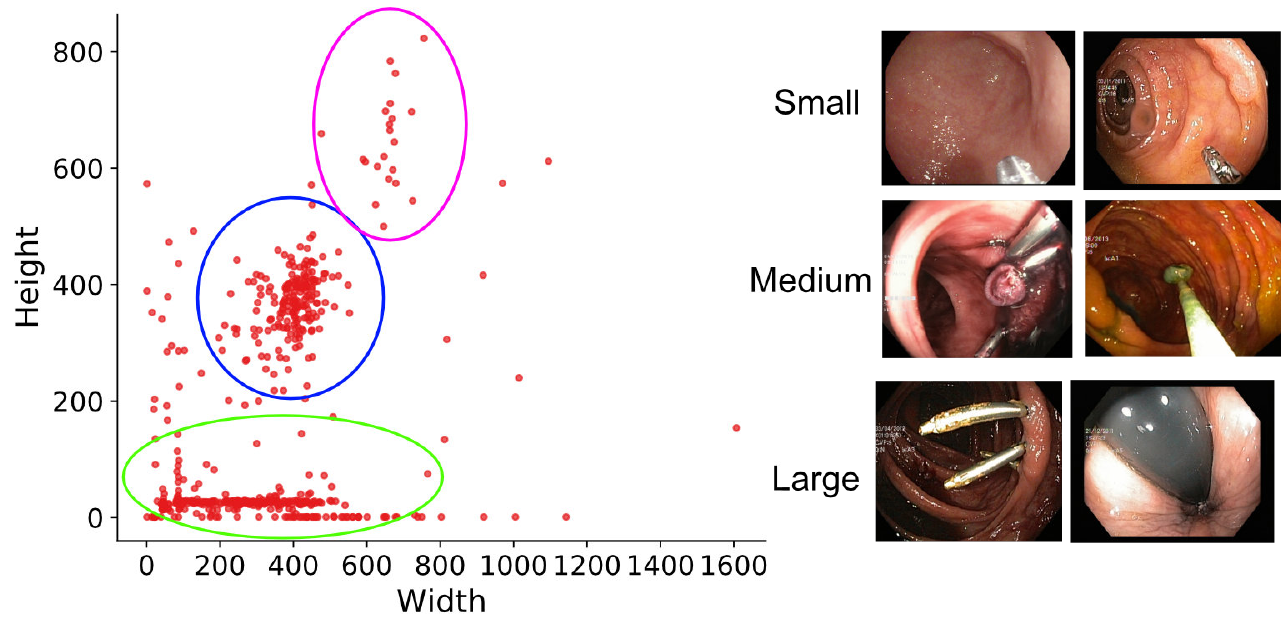}
    \caption{Distribution of Kvasir-Instrument dataset. On left: Small (green), medium (blue) and large (pink) sized tool clusters. On right: sample images with variable tool size in images.}
    \label{fig:distribution}
\end{figure}

\begin{figure}[t!]
    \centering
    \includegraphics[width= 1.0\textwidth]{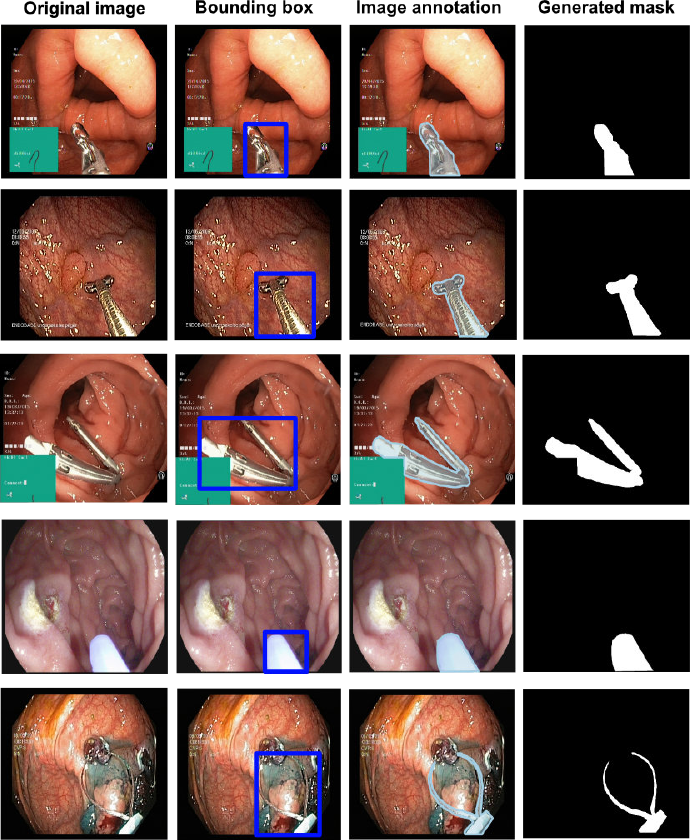}
    \caption{\textbf{Kvasir-Instrument dataset:} First two rows represent frames with biopsy forceps, the middle row consist of metallic clip, the fourth row is a radio-frequency ablation probe and the last row depicts the crescent and hexagonal shaped snares for polyp removal.}
    \label{fig:datasetsample}
\end{figure}

In this section, we introduce the Kvasir-Instrument dataset with details on how the data was collected, the annotation protocol, and the dataset's structure. The dataset was collected from endoscopic examinations performed at the Bærum Hospital in Norway. The unlabelled images' frames are selected from the HyperKvasir dataset~\cite{hyperkvasir2020}. 
%The Hyperkvasir dataset contains a total of 110,079 images and 373 videos, out of which only 10,662 are labeled by the endoscopists. However, there still exist 99,417 unlabeled images from the different parts of the \ac{GI} tract. 
%This suggests that there is still an opportunity for labeling the remaining classes. The labeling of medical dataset is a manual process. It requires effort from expert endoscopists to label this large number of images, which is both time-intensive and costly. 
%
HyperKvasir provides frame-level annotations for 10,662 frames for 23 different classes. However, the majority of the images (99,417 frames) are not labeled. We trained a model using the labeled samples of this dataset and tried to predict the classes of the unlabeled samples. Although our algorithm~\cite{thambawita2018medico,thambawita2020extensive} could not classify all the images correctly; however, we were able to classify the instrument class out of hundreds of thousands of provided image frames. Additionally, some images were extracted manually from the polyp class of the Kvasir-SEG~\cite{jha2020kvasir} dataset. Below, we present the acquisition and annotation protocols used in the data preparation:
%
%\subsection{Collection protocol} 
\paragraph{Data acquisition:}
The images and videos were collected using standard endoscopy equipment from Olympus (Olympus Europe, Germany) and Pentax (Pentax Medical Europe, Germany) at Vestre Viken Hospital Trust, Norway. All the data used in this study were obtained from videos for procedures that had followed the patient consenting protocol of Bærum Hospital. Additionally, no patient information was used for archiving. We have performed a random naming for each publicly released frame for effective annonymisation. 

%\subsection{Annotation protocol}
\paragraph{Annotation strategy:}
We have uploaded the Kvasir-Instrument dataset to  labelbox\footnote[5]{\url{https://www.labelbox.com/}} and labeled the \ac{ROI} in the image frames, i.e., the ROI of diagnostic and therapeutic tools in our cases and generated all the ground truth masks. Figure~\ref{fig:datasetsample} shows the example images, bounding box, image annotation, and generated masks for the Kvasir-Instrument dataset. All annotations were then exported in a JSON format which was used to generate masks for each of the annotations. Related codes and more information about the dataset can be found here\footnote[6]{\url{https://github.com/DebeshJha/Kvasir-Instrument}}.

% TODO: make a github code repo and put the extraction code file (if doable!)
The exported file contained the information of the images along with the coordinate points that were used for mask and bounding box generation. All annotations were performed using a three-step strategy:
\begin{itemize}
    \item First, the selected samples were labeled by two experienced research assistants.
    \item The annotated samples where cross-validated for their delineation quality by two experienced GI experts (more than 10 years of work experience in colonoscopy).
    \item Finally, the suggested changes were incorporated using the comments from experts and were validated for only those samples.
\end{itemize}

%\paragraph{Kvasir-Instrument dataset:}
The Kvasir-Instrument dataset includes 590 frames consisting of various GI endoscopy tools used during both, endoscopic surveillance and therapeutic or surgical procedures. A thorough annotation strategy (detailed above) was used to create bounding boxes and segmentation masks. The dataset consists of variable tool size with respect to image height and width as presented in Figure~\ref{fig:distribution}. The majority of the tools  are small and medium-sized. The sample bounding box annotation, precise area delineation and extracted masks, are shown in Figure~\ref{fig:datasetsample}.

%TODO!!! 
Our dataset is publicly available, and can be accessed at: \url{https://datasets.simula.no/kvasir-instrument/}. It consists of original image samples (in JPEG format), their corresponding masks (in PNG format), and bounding box information (in JSON format). A sample python script to help researchers visualise the data is also provided. 
\section{Benchmarking, results and discussion}
In this section, we explore encoder-decoder based classical models for baseline algorithm benchmarking, their implementation details for reproducibility, details on evaluation metric used for quantitative analysis, and results and discussion. 
\subsection{Baseline methods}
U-Net has been explored in the past through many biomedical segmentation challenges and has shown strength towards an effective supervised segmentation model. In this paper, we, therefore, use U-Net based architectures on our Kvasir-Instrument dataset to provide a baseline result for future comparisons. U-Net uses an encoder-decoder architecture, that is, a contractive feature extraction path and expansive path with a classifier to perform  binary classification of each image pixel in an upsampled feature map.
In our previous work, we have shown that the strength of supervised classification can be amplified by using the output mask from one U-Net~\cite{ronneberger2015u} architecture to the other by proposing DoubleUNet~\cite{jha2020doubleu}. In addition, the DoubleUNet architecture uses VGG-19 pretrained on ImageNet as one of the encoder block, squeeze and excite block and \ac{ASPP} block. All other components in the network remain the same as the U-Net. 
For both networks, dice loss gives an $1-DSC$, where DSC is the dice similarity coefficient (see Eq.~\ref{eq:1} below).

\subsection{Implementation Details}
We have implemented the U-Net-based and DoubleUNet based architectures using the Keras framework~\cite{chollet2015keras} with TensorFlow~\cite{abadi2016tensorflow} as backend running on the Experimental Infrastructure for Exploration of Exascale Computing (eX3), NVIDIA DGX-2 machine. We have resized the training dataset into 512$\times$512. We set the batch size of 8 for training. Both architectures are optimized by using the Adam optimizer. We have made use of dice loss as the loss function. We split the dataset using 80\% of the dataset for training and the remaining 20\% for the testing (evaluation). We performed basic augmentation, such as horizontal flip, vertical flip, and random rotation. Moreover, we have also provided the train-test split so that others can improve the methods on the same dataset. 
%can we share the pertained models too? we should also mention that the train and test set split for the experiments is also shared
\subsection{Evaluation Metrics}
\label{sec:Suggested_Metrics}
In this medical image segmentation approach, each pixel of the diagnostic and therapeutic tool either belongs to a tool  or non-tool region. Dice similarity coefficient (DSC) is the main evaluation metric used to evaluate this task. Additionally, we calculate other standard metrics such as Jaccard similarity coefficient (JC) or intersection over union (IoU), precision, recall, overall accuracy, F2, and frames per second (FPS) as it is a commonly used metric in biomedical image segmentation tasks. The mathematical expressions for them are as follows:
\begin{equation}{\label{eq:1}}
\text{DSC} = \frac{2 \cdot tp} {2 \cdot tp + fp + fn}
\end{equation}
\begin{equation}{\label{eq:2}}
\text{JC or IoU} = \frac{tp} {tp + fp + fn}
\end{equation}
\begin{equation}{\label{eq:3}}
\text{Recall}~(r) = \frac{tp} {tp + fn}
\end{equation}

\begin{equation}{\label{eq:4}}
\text{Precision}~(p) =\frac{tp} {tp + fp}
\end{equation}
\begin{equation}{\label{eq:5}}
\text{F2} = \frac{5p \times r} {4p + r}
\end{equation}
\begin{equation}{\label{eq:6}}
\text{Overall accuracy}~(Acc.) ={\frac{tp + tn} {tp + tn + fp + fn}} 
\end{equation}

\begin{equation}{\label{eq:7}}
\text {Frame Per Second}~(FPS) = {\frac{1} {sec /frame}}
\end{equation}
%%%%%%%%%%%

Here, tp, fp,  tn, fn are the true positives, false positive, true negative, and false negative, respectively. 

\begin{table*}[t!]
\footnotesize
\centering
\def\arraystretch{1.2}
\caption{Baseline results for tool segmentation}% on the Kvasir-Instrument dataset}
\label{res:polypsegmentation}
\begin{tabular}{l|c|c|c|c|c|c|c}
\toprule
\textbf{Method} & \textbf{JC}  & \textbf{DSC} & \textbf{F2-score} & \textbf{Precision} & \textbf{Recall} & \textbf{ Acc.} & \textbf{FPS} \\ \midrule
{U-Net}~\cite{ronneberger2015u} &   \textbf{0.8578 }  &  \textbf{0.9158}   & \textbf{0.9320}  & \textbf{0.8998} &\textbf{0.9487 }  &\textbf{0.9864} & \textbf{20.4636}\\ \hline
{DoubleUNet}~\cite{jha2020doubleu} & 0.8430    & 0.9038 &  0.9147  &  0.8966    & 0.9275    &  0.9838 & 10.0000  \\
\bottomrule
\end{tabular}
\end{table*}

\begin{figure}[t!]
    \centering
    \includegraphics[width= 1.0\textwidth]{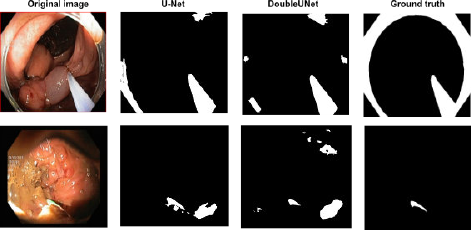}
    \caption{\textbf{Failed cases:} Cap region (top) is under-segmented and small clip area is over-segmented and consist of large number of false positives (bottom).}
    \label{fig:failed_cases}
\end{figure}
\subsection {Quantitative and Qualitative results}
Table~\ref{res:polypsegmentation} shows the results of the baseline methods for the tool segmentation on the Kvasir-Instrument dataset. From the table, we can observe that the UNet achieved a high JC of 0.8578 and DSC of 0.9158, which is slightly above than the DoubleUNet that yielded JC of 0.8430 and DSC of 0.9038. Also, UNet achieved a speed of 20.4636 FPS, whereas computational time is double for DoubleUNet with only 10 FPS. Similarly, both the recall and precision scores are very comparable for both U-Net ($p = 0.8998, r = 0.9487$) and DoubleUNet ($p = 0.8966, r = 0.9275$).
%could only achieve an FPS of 10 FPS. The highest scores are bold.  The highest scores are bold. 

Figure~\ref{fig:failed_cases} shows the qualitative result on the challenging images. It can be observed that that both UNet and DoubleUNet are under-segmenting the cap region (top) and over-segmenting the small clip area (bottom). Some parts of these images are confused because of the presence of saturation areas. However, both models was able to segment well with most endoscopic tool samples in the dataset. This is also evident from the quantitative results. 

\subsection{Discussion}
\label{sec:discussion}
From the experimental results in Table~\ref{res:polypsegmentation}, we can validate that the classiccal U-Net architecture outperforms DoubleUNet model. Additionally, U-Net is $2 \times$ faster than the DoubleUNet. This is because U-Net uses basic convolution blocks, whereas DoubleUNet uses pre-trained encoders, \ac{ASPP}, squeeze and excite blocks, all of which increases the inference latency. Here, the UNet is optimized by dice loss instead of binary cross-entropy loss, which showed improved performance during our experiments.

Further, fine-tuning on other similar datasets, rigorous data augmentation and applying more advanced \ac{DL} techniques can improve the baseline results - eventually achieving the detection, localisation, and segmentation performance needed to make the technology useful in a clinical environment. Additionally, use of \ac{DL} networks with less parameters could increase the computational efficiency thereby enabling real-time systems that can be used in clinical settings effectively. 

\section{Conclusion}
\label{sec:conclusion}
We have curated, annotated, and publicly released a dataset that incorporates tools used in GI endoscopy screening and surgical procedures. The dataset consists of images, bounding boxes and segmentation masks of endoscopy tools used during different procedures in the GI tract. Additionally, we provided baseline segmentation methods for the automatic delineation of these tools and have compared them using standard computer vision metrics. In the future, we plan to continuously increase the amount of data and also call for multi-media challenges on using the presented dataset. 

\section*{Acknowledgements}
This work is funded in part by the Research Council of Norway, project number 263248 (Privaton) and project number 282315 (AutoCap). We performed all computations in this paper on equipment provided by the Experimental Infrastructure for Exploration of Exascale Computing ($eX^3$), which is financially supported by the Research Council of Norway under contract 270053.

\bibliographystyle{splncsnat}
\bibliography{references.bib}
\end{document}